\documentclass[aps,prd,a4paper,onecolumn,amsmath,showpacs,superscriptaddress,nofootinbib,preprintnumbers,notitlepage]{revtex4-1}
\usepackage{amssymb,amsmath}
\usepackage{graphicx}
\usepackage{color}
\usepackage{pgfplots}
\usepackage{booktabs}
\usepackage{multirow}
\usepackage{dcolumn}
\usepackage{amsmath}
\usepackage{mathtools}
\usepackage{amsfonts}
\usepackage{amssymb}
\usepackage{epstopdf}
\usepackage{bm}
\usepackage{braket}
\usepackage{enumitem}
\usepackage{soul}
\usepackage{ulem}
\usepackage{color}
\usepackage{transparent}
\usepackage{pifont}
\usepackage[font={small}]{caption}
\usepackage{float}
\usepackage{orcidlink}

\pgfplotsset{compat=1.18}

\usepackage{mathrsfs}
\begin{document}

\title{Reconstructing inflation in a generalized Rastall theory of gravity}

\author{Ram\'on Herrera\orcidlink{0000-0002-6841-1629}}
\email{ramon.herrera@pucv.cl}
\affiliation{Instituto de F\'{\i}sica, Pontificia Universidad Cat\'{o}lica de Valpara\'{\i}so, Avenida Brasil 2950, Casilla 4059, Valpara\'{\i}so, Chile.
}

\author{Carlos Ríos\orcidlink{0000-0003-3228-0003}}
\email{carlos.rios@ucn.cl}
\affiliation{Departamento de Ense\~nanza de las Ciencias B\'asicas, Universidad Cat\'olica del Norte, Larrondo 1281, Coquimbo, Chile.}

\begin{abstract}
We investigate the reconstruction of standard and generalized Rastall gravity inflationary models, using the scalar spectral index and the Rastall parameter expressed as functions of the number of $e-$folds $N$. Within a general formalism, we derive the effective potential in terms of the relevant cosmological parameters and the Rastall parameter for these gravity frameworks. As a specific example, we analyze the attractor $n_s(N) - 1 \propto N^{-1}$, first by considering constant values of the Rastall parameter to reconstruct the inflationary stage in standard Rastall gravity, and then by assuming a linear dependence on the number of $e-$folds $N$ to reconstruct the inflationary model in generalized Rastall gravity. Thus, the reconstruction of the potential $V(\phi)$ is obtained for both standard and generalized Rastall gravity inflationary models. In both frameworks, we constrain  key parameters of the reconstructed models during inflation using the latest observational data from Planck.
\end{abstract}

\maketitle

\section{Introduction}

According to the theory of cosmic inflation, the early universe underwent a brief phase characterized by accelerated expansion  \cite{Starobinsky:1980te,Guth:1980zm}. Originally proposed to address several unresolved issues within the standard hot Big Bang paradigm, such as the horizon, flatness, and monopole problems, inflation also offers a compelling mechanism for the generation of primordial density perturbations. These perturbations are the seeds of the large-scale structure (LSS) observed today \cite{Mukhanov:1981xt,Hawking:1982cz}, and they account for the anisotropies detected in cosmic microwave background radiation (CMB) \cite{Guth:1982ec}.

Among the various inflationary scenarios developed to explain the dynamics of the early universe, considerable attention has been given to models that extend Einstein’s general relativity. These extensions often involve generalized Lagrangian densities or actions that go beyond the standard Einstein–Hilbert action, thereby modifying the equations of motion and, consequently, the inflationary phase. However, an alternative approach to modifying gravitational theory is through a reformulation of the conservation law of the energy–momentum tensor. Unlike general relativity, where the energy–momentum tensor is covariantly conserved, these models typically allow for a non-vanishing divergence while still preserving a minimal coupling between matter and spacetime geometry.

Rastall gravity, introduced in Ref.\cite{Rastall:1972swe}, is a notable example of a modified theory in which the energy–momentum tensor is not covariantly conserved.  This theory challenges the universal validity of the covariant conservation of the energy–momentum tensor, noting that this condition is typically derived under specific assumptions, such as flat Minkowski spacetime or weak gravitational fields. Rastall argued that, in strongly curved spacetimes like those present in the early universe or near compact objects, the conservation law might no longer hold in its standard form. Following Rastall's proposal, the standard assumption of a vanishing covariant divergence of the energy-momentum tensor is relaxed. In this framework, the non-conservation is assumed to be proportional to the gradient of the Ricci scalar $R$, such that $\nabla_\mu T^{\mu\nu}\propto \nabla^{\nu} R$, where $T^{\mu\nu}$ denotes  the energy-momentum tensor. One of the main limitations of Rastall's theory lies in its phenomenological character and the lack of a fundamental action principle from which it can be consistently derived. Despite this, the theory exhibits a rich structure that connects to several foundational aspects of gravitational physics \cite{Birrell:1982ix}. In particular, Rastall's theory can be interpreted as a phenomenological framework for capturing the covariant effects of quantum fields in curved spacetime \cite{Bertlmann:1996xk}. Although the theory lacks a fundamental action formulation, it is possible to construct an effective action by introducing an external field into the Einstein–Hilbert action, typically through the use of a Lagrange multiplier or a particular case of $f(R,T)$ gravity, with $T$ the trace of the energy-momentum tensor \cite{DeMoraes:2019mef,Shabani:2020wja}. However, as demonstrated in Ref.\cite{Fabris:2020uey}, the original formulation of Rastall gravity presents significant obstacles to its embedding within a conventional Lagrangian approach. In relation to the cosmology to describe the universe, 
different studies have investigated Rastall gravity as a framework for both early-universe inflation and late-time acceleration. In relation to the inflationary models in this theory, typically involve scalar fields whose dynamics are altered by the non-conservation of the energy-momentum tensor, allowing compatibility with observational bounds on the spectral index and tensor-to-scalar ratio, see e.g.,  \cite{Batista:2011nu,Afshar:2023uyw,Mandal:2023ink}.
At late times, Rastall gravity can reproduce dark energy behavior in order to describe the dark sector at the present. In this stage, solutions involving barotropic fluids or scalar fields can mimic quintessence or phantom regimes depending on the Rastall parameter, see   Refs.\cite{Fabris:2017msx,Singh:2024urv,Singh:2024urv}. In addition, in Ref.\cite{Moradpour:2017shy}, an extension of Rastall gravity, referred to as generalized Rastall gravity, was introduced, where the Rastall parameter is no longer assumed to be constant but is instead allowed to evolve dynamically throughout cosmic history. In the context of this generalized framework, the non-conservation is assumed to be proportional to the gradient of the product of the Rastall parameter and the Ricci scalar. Here, the authors show that the model exhibits a nonsingular emergent scenario during the initial stage of cosmic expansion for a specific choice of the Rastall parameter. Furthermore, it is shown to be equivalent to the particle creation mechanism in Einstein gravity within the framework of nonequilibrium thermodynamics. In Ref.\cite{Das:2018dzp}, building on this modification of Rastall gravity, the authors investigated the evolution of both linear and nonlinear matter perturbations in the universe by employing the spherically symmetric top-hat collapse framework. Thus, several cosmological models have been investigated in the literature within the framework of a generalized Rastall theory of gravitation, in which the Rastall parameter is treated as a variable (see, e.g., Refs.\cite{Lin:2018dgx,Lin:2020fue,Ziaie:2020ord,Shabani:2022zlx}).

On the other hand, various authors have explored the reconstruction of the effective inflationary potential in the inflationary epoch by utilizing observational quantities such as the scalar power spectrum, the scalar spectral index, and the tensor-to-scalar ratio, see Refs.\cite{Hodges:1990bf,Easther:1995pc,Martin:2000ei,Herrera:2015udk,Belinchon:2019nhg}. Reference \cite{Hodges:1990bf} was the first to explore the reconstruction of inflationary potentials in the framework of a single-scalar field, based on the primordial scalar power spectrum. This analysis was carried out within the general slow-roll approximation, without assuming a particular form for the scalar spectral index. An effective methodology for reconstructing the inflationary potential within the slow-roll approximation consists of parameterizing the scalar spectral index (or cosmological attractor)  
$n_s=n_s(N)$, where
$N$ denotes the number of $e-$folds during the inflationary epoch. Regarding the parametrization of the scalar spectral index as a function of the number of 
$e-$folds, the constraints obtained from the Planck observational data \cite{Planck2018} show strong agreement with the expression
$n_s(N)\sim 1-2/N$, which emerges in a variety of inflationary models under the slow-roll approximation, assuming large $N$, see e.g., $T$ and $E$ models \cite{Kallosh:2013hoa,Kallosh:2013maa}, $R^2-$model \cite{Starobinsky:1980te}, chaotic model \cite{Linde:1983gd} and  among other. This agreement is valid under the common assumption that the number of 
$e-$folds occurring between the horizon exit of the cosmological scales and the end of inflation lies approximately within the range
$N\sim 55-70$. This reconstruction methodology based on the attractor $n_s(N)$ has also been employed to investigate different inflationary cosmological models; see Refs.\cite{Herrera:2018cgi,Herrera:2018mvo,Herrera:2019xhs,Herrera:2024aqf,Herrera:2024ojo,Herrera:2022kes}.
Alternatively, by parametrizing the slow roll parameter
$\epsilon=\epsilon(N)$ as a function of
$N$, one can reconstruct the effective potential, the scalar spectral index and the tensor-to-scalar ratio and subsequently compare these results with observational data \cite{Mukhanov:2013tua,Huang:2007qz,Lin:2015fqa}. In addition, the author in Ref.\cite{Roest:2013fha} extends this approach using both slow-roll parameters; $\epsilon(N)$ and $\eta(N)$, see also Refs.\cite{Garcia-Bellido:2014gna,Creminelli:2014nqa}. However, these methodologies do not necessarily ensure consistency with the observational constraints on the cosmological parameters $n_s$ and $r$ within the range  $N\sim 55-70$.

This work aims to reconstruct a Rastall inflationary model by adopting a parameterization of the scalar spectral index as a function of the number of
$e$-folds. Within this framework, we investigate how the background dynamics influences the reconstruction of the effective potential associated with this attractor. In this study, we initially reconstruct the standard Rastall gravity, where the Rastall parameter is considered as a constant. Subsequently, we explore a generalized formalism of Rastall gravity, characterized by an extended Rastall framework, to reconstruct the inflationary scenario.
In both frameworks, by adopting a general formalism, we derive the inflationary potential under the slow roll approximation, based on the Rastall parameter together with the attractor associated to the scalar spectral index $n_s(N)$ and expressed in integral form.
To reconstruct analytical expressions for the potential, we study a specific example of the cosmological parameter associated with the scalar spectral index and the Rastall parameter. In this context, we consider the attractors $n_s - 1 \propto 1/N$ in two frameworks during the stages of Rastall inflation. In both scenarios, we reconstruct the potential as a function of the number of $e$-folds and in terms of the scalar field, while also deriving constraints on the model parameters under the assumption of the observational data.

The outline of the article is as follows: the next section presents a short review of the basic equations during the stage of Rastall gravity. In this section, we present the fundamental relationships within the framework of standard Rastall gravity, where the Rastall parameter is taken as constant. In Section II-A,  we obtain, under a general formalism, explicit expressions for the effective potential  in terms of the number of $e-$folds $N$ during the stage of the standard Rastall theory. In Section II-B, we discuss a concrete example for our model, in which we consider the specific attractors for $n_s(N)$, in order to reconstruct the potential $V(\phi)$. In Section \ref{Gen} we study the general formalism of the reconstruction of inflation in the framework of a generalized Rastall gravity, where the Rastall parameter is no longer treated as a constant. In Section III-A, we analyze a specific example for the Rastall parameter and the scalar spectral index in terms of the number of $e-$ folds $N$ to reconstruct the effective potential as a function of the scalar field.  In Section \ref{SA}, we study the stability analysis for our reconstructed models in the framework of standard and generalized Rastall gravity.
Finally, our conclusions are presented in Sect.\ref{Conc}. We chose units such that
$c=\hbar=1$.

\section{Rastall Gravity}\label{RGI}

In this section, we present a brief overview of Rastall's theory proposed in Ref.\cite{Rastall:1972swe}.  As is well known, Rastall theory introduces a modification to the Einstein field equations in which the energy–momentum tensor $T_{\mu\nu}$ is not covariantly conserved.
In this framework, the gravitational field equations take the form \cite{Vitohekpon:2025nzq}

\begin{equation}
\label{campoRas}
    G_{\mu\nu} +\frac{1}{2}\left(1-\lambda_{\text{Ras}}\right) g_{\mu\nu} R =R_{\mu\nu}-\frac{1}{2}\lambda_{\text{Ras}}\,g_{\mu\nu}\,R= \kappa T_{\mu\nu},
\end{equation}
where the quantity $\lambda_\text{Ras}$, known as the Rastall parameter, is taken as constant in the framework of standard Rastall gravity and as variable in the generalized Rastall theory of gravity. Moreover, $\kappa=\alpha\kappa^\text{(GR)}$ denotes the Rastall gravitational constant, with $\kappa^\text{(GR)}=8\pi G=8\pi/M_p^2$ the gravitational coupling in GR, where $M_p$ is the Planck mass. Additionally, $R$ denotes the Ricci scalar, the tensor $G_{\mu\nu}=R_{\mu\nu}-(1/2)g_{\mu\nu}R$, and the quantity $g_{\mu\nu}$ corresponds to the metric tensor.

In relation to the parameter $\alpha$ and following Ref.\cite{Rastall:1972swe}, it is defined in terms of $\lambda_\text{Ras}$, which arises from the Newtonian limit, as
\begin{equation}
\alpha=\frac{(2\lambda_\text{Ras}-1)}{(3\lambda_\text{Ras}-2)},\,\,\,\,\mbox{with}\,\,\,\,\,\lambda_\text{Ras}\neq \frac{2}{3}.
\end{equation}
Thus, the Rastall gravitational constant $\kappa$ is given by 
\begin{equation}
\kappa=\left(\frac{8\pi}{M_p^2}\right)\,\,\left[\frac{(2\lambda_\text{Ras}-1)}{(3\lambda_\text{Ras}-2)}\right]=\kappa^\text{(GR)}\,\alpha.\label{kras}
\end{equation}
In the particular case where the Rastall parameter $\lambda_{\text{Ras}} = 1$, the Rastall gravitational constant reduces to the standard gravitational constant of GR, and Eq.(\ref{campoRas}) becomes the standard Einstein field equations.

By applying the covariant derivative to both sides of the Eq.(\ref{campoRas}), and utilizing the Bianchi identity $\nabla^\mu G_{\mu\nu}=0$, it follows that the covariant divergence of the energy–momentum tensor takes the form of
\begin{equation}
\label{noncT}
\nabla_\mu T^{\mu\nu}= \frac{1-\lambda_\text{Ras}}{2\kappa}\nabla^\nu R\propto \nabla^\nu R,
\end{equation}
and it corresponds to the Rastall's original field equations \cite{Rastall:1972swe}.
In addition, from the above equation, one can see that energy-momentum conservation is restored for $\lambda_{\text{Ras}} = 1$, such that $\nabla_\mu T^{\mu\nu} = 0$, as in the standard formulation of GR.
Nevertheless, Eq.(\ref{campoRas}) can be rewritten in the conventional form of Einstein’s field equations by defining an effective energy-momentum tensor, $T_{\mu\nu}^{\text{eff}}$, as follows
\begin{equation}
\label{Teff}
T_{\mu\nu}^{eff}=T_{\mu\nu} - \frac{1-\lambda_{\text{Ras}}}{2(1-2\lambda_{\text{Ras}})}g_{\mu\nu}T,\,\,\,\,\,\,\,\,\,\mbox{with}\,\,\,\,\,\,\,\,\lambda_\text{Ras}\neq \frac{1}{2}.\end{equation}
Here, $T=g^{\mu\nu}T_{\mu\nu}$ represents the trace of the momentum energy tensor. In this form, we can rewrite  Eq.(\ref{campoRas}) as
\begin{equation}
    G_{\mu\nu}= R_{\mu\nu} - \frac{1}{2}g_{\mu\nu} R= \kappa T_{\mu\nu}^{eff}.\label{G}
\end{equation}
An important observation that arises from considering the Newtonian limit is the constraint on the Rastall gravitational constant, along with the implication for the effective energy-momentum tensor, in which the Rastall parameter
$\lambda_\text{Ras}$ becomes singular for
$\lambda_\text{Ras} \neq 2/3$ and $\lambda_\text{Ras} \neq 1/2$, respectively. Such divergences indicate unphysical behavior, suggesting that these values of
$\lambda$ lie outside the physically acceptable range; see, for example, Refs.\cite{Moradpour:2016ubd,Hansraj:2018zwl}.

\subsection*{Standard Rastall gravity: basic relations}\label{CaseI}

We begin by investigating the inflationary reconstruction and dynamics of the early universe in the framework of standard Rastall gravity in which the Rastall parameter $\lambda_\text{Ras}$ is a constant.  Before proceeding,  it is essential to specify the nature of the matter responsible for driving inflation. In this sense, we assume that the matter is described by a canonical scalar field $\phi$. This choice enables us to explore the role of the scalar field as the source of matter driving the early cosmological dynamics within a non-conservative gravitational framework.

In this context, the Lagrangian density related to scalar field $\phi$ is defined as
\begin{equation}
    \mathcal{L}=\frac{1}{2}g^{\mu\nu}\partial_\mu\phi\partial_\nu\phi-V(\phi)=X-V(\phi),
\end{equation}
where the first term $X=\frac{1}{2}g^{\mu\nu}\partial_\mu\phi\partial_\nu\phi$ represents the kinetic energy of the scalar field and $V(\phi)$ denotes the effective potential associated with the scalar field. In this sense, the energy-momentum tensor of the scalar field $T_{\mu\nu}$, is defined as
\begin{equation}
    \label{thetamunu}T_{\mu\nu}=-\frac{2}{\sqrt{-g}}\frac{\partial\left(\sqrt{-g}\mathcal{L}\right)}{\partial g^{\mu\nu}}=\partial_\mu\phi\partial_\nu\phi-g_{\mu\nu}\left(\frac{1}{2}\partial^\alpha\phi\partial_\alpha\phi+V(\phi)\right).
\end{equation}
 The effective energy-momentum tensor given by Eq.(\ref{Teff}), can be recast into the form of a perfect fluid energy-momentum tensor. In this way, 
by employing Eqs.(\ref{Teff}) and (\ref{thetamunu}), we can obtain the expressions for  the energy density $\rho=T^{eff}_{00}$ and the pressure $pg_{ij}=T^{eff}_{ij}$, associated with the scalar field and these quantities are given by
\begin{eqnarray}
    \rho&=&\frac{1}{2\lambda_\text{Ras}-1}\left[\frac{3\lambda_\text{Ras}-2}{2}\dot{\phi}^2+V(\phi)\right]=\frac{(3\lambda_\text{Ras}-2)}{(2\lambda_\text{Ras}-1)}\,X+\frac{V}{2\lambda_\text{Ras}-1},\,\,\,\,\,\,\,\,\,\,\,\mbox{and}\label{rho}\\
    p&=&\frac{1}{2\lambda_\text{Ras}-1}\left[\frac{\lambda_\text{Ras}}{2}\dot{\phi}^2-V(\phi)\right]=\frac{\lambda_\text{Ras}}{(2\lambda_\text{Ras}-1)}\,X+\frac{V}{2\lambda_\text{Ras}-1}\label{p},
\end{eqnarray}
respectively. Here we have assumed that the scalar field corresponds to a homogeneous scalar field such that $\phi=\phi(t)$ and we have also used the definition of the effective energy-momentum tensor $T^{eff}_{\mu\nu}$ given by Eq.(\ref{Teff}).
In addition, in the following we will consider that the dots denote the derivative with respect to cosmic time $t$.

On the other hand, considering a spatially flat Friedmann-Robertson-Walker (FRW) metric, the Friedmann equation from Eq.(\ref{G}) takes the form of
\begin{equation}
    H^2 =\frac{\kappa}{3}\,\rho= \frac{\kappa}{3(2\lambda_\text{Ras} - 1)}\left[\frac{(3\lambda_\text{Ras} - 2)}{2}\dot{\phi}^2+V(\phi)\right],\label{FIcom}
\end{equation}
where $H=\dot{a}/a$ corresponds to the Hubble parameter and $a$ is the scale factor.

Applying the modified continuity equation given by $\dot{\rho}+3H(\rho+p)=0$, we derive the equation of motion that governs the scalar field within the framework of a standard Rastall theory. Thus, the dynamics of the scalar field is governed by the corresponding Klein–Gordon equation, which takes the form of
\begin{equation}
(3\lambda_\text{Ras}-2)\ddot{\phi}+3H(2\lambda_\text{Ras}-1)\dot{\phi}+V_\phi=0.\label{KGcom1}
\end{equation}
In the following,  the notation $V_\phi$ corresponds to $V_\phi=\partial V/\partial \phi$, $V_{\phi\phi}$ to $V_{\phi\phi}=\partial^2 V/\partial \phi^2$, etc.

During  the inflationary era, we consider the slow-roll approximation, under which the kinetic energy of the scalar field is significantly smaller than its potential energy, and the acceleration term associated with the scalar field in the equation of motion can also be neglected. Under this approach, the Friedmann equation (\ref{FIcom}) and the Klein–Gordon equation (\ref{KGcom1}) become
\begin{equation}
    H^2=\frac{\kappa}{3}\,\rho \simeq \frac{\kappa}{3(2\lambda_\text{Ras} -1)}V(\phi),\,\,\,\,\,\,\,\mbox{and}\,\,\,\,\,\,\,\,\,\,\,\,3H(2\lambda_\text{Ras}-1)\dot{\phi}+V_\phi\simeq0,
    \label{EqsinSR}
\end{equation}
respectively.

Furthermore, we introduce the so-called slow-roll parameters, $\epsilon$ and $\eta$, defined as follows: 
\begin{equation}
    \epsilon=-\frac{\dot{H}}{H^2},\,\,\,\,\,\,\,\mbox{and}\,\,\,\,\,\,\,\,\,
    \eta=-\frac{1}{H}\frac{\ddot{\phi}}{\dot{\phi}}\label{etadef}.
\end{equation}
A necessary condition for inflation to occur is that the slow-roll parameter satisfies the inequality $\epsilon < 1$. which is analogous to $\ddot{a}>0$ (from Eq.(\ref{etadef})). The inflationary epoch ends when the parameter reaches $\epsilon = 1$, or equivalently, when $\ddot{a} = 0$.

 The slow roll parameters $\epsilon$ and $\eta$ can also be expressed as conditions on the shape of the effective potential. In this way, using Eqs.(\ref{EqsinSR}) and (\ref{etadef}), we can write the slow-roll parameter $\epsilon$ as
\begin{equation}
    \epsilon_V=\frac{3(2\lambda_\text{Ras}-1)}{2V}\dot{\phi}^2=\frac{1}{2\kappa}\left(\frac{V_\phi}{V}\right)^2=\alpha^{-1}\epsilon_V^\text{(GR)},\label{p1}
\end{equation}
where the subindex $V$ denotes that the  parameter $\epsilon$ depends on the form of the potential.  For the second slow-roll parameter $\eta$, we find that it can be written as follows 
\begin{equation}
\eta=\frac{1}{3(2\lambda_\text{Ras}-1)}\left(\frac{V_{\phi\phi}}{H^2}+\frac{V_\phi}{H\dot{\phi}}\epsilon\right)=\frac{1}{\kappa}\frac{V_{\phi\phi}}{V}-\epsilon_V=\eta_V-\epsilon_V=\alpha^{-1}\left(\eta_V^\text{(GR)}-\epsilon_V^\text{(GR)}\right), \label{p2}
\,\,\,\,\,\,\mbox{with}
\,\,\,\,\,\,\,
\eta_V=\frac{1}{\kappa}\frac{V_{\phi\phi}}{V}.
\end{equation}
Here, we observe that the slow-roll parameters given by Eqs.(\ref{p1}) and (\ref{p2}), associated with Rastall's theory, reduce to the standard slow-roll parameters of GR when $\alpha = 1$. 

Furthermore, in order to determine the duration of the inflationary period, it is useful to introduce the number of $e-$folds inflation, denoted by $N$. In this context, the number of $e-$folds between two different values of cosmological times $t_\text{end}$ and $t$ (or $\phi_\text{end}$ and $\phi$ )is defined as
\begin{equation}
\label{deltaN}
    \Delta N=N-N_\text{end}=\ln\left[\frac{a(t_\text{end})}{a(t)}\right]=\int^{t_\text{end}}_t H\,dt=\int^{\phi_\text{end}}_\phi\left(\frac{H}{\dot{\phi}}\right)d\phi,
\end{equation}
where $N_\text{end}$ and $t_\text{end}$ (or $\phi_\text{end}$) represent the values of the number of $e-$folds and the time (or scalar field) at which the inflationary phase ends, respectively. Here, we have used the fact that the relation between $dN$ and $dt$ is given by $dN=-H\,dt$, see, e.g. \cite{Chiba:2015zpa,Herrera:2015udk}.  

From the energy density and pressure defined by Eqs.(\ref{rho}) and (\ref{p}), we observe that the kinetic terms exhibit couplings described by the Rastall parameter. This suggests that the adiabatic sound speed $c_s$ differs from unity and depends on this parameter, since it is defined as \cite{Garriga:1999vw,DeFelice:2013ar}
\begin{equation}
c_s^2=\frac{(\partial p/\partial X)}{(\partial \rho/\partial X)}=\frac{p_X}{\rho_X}.
\end{equation}
In this way, from Eqs.(\ref{rho}) and (\ref{p}) we find that the adiabatic speed of sound squared in the standard Rastall gravity, where the parameter $\lambda_\text{Ras}$ is taken to be constant, results in
\begin{equation}
c_s^2=\frac{p_X}{\rho_X}=\frac{\lambda_\text{Ras}}{(3\lambda_\text{Ras}-2)}=\mbox{constant}.\label{19}
\end{equation}
Here, we note that for the specific value in which the Rastall parameter $\lambda_\text{Ras}\rightarrow 1$, the sound of speed given by Eq.(\ref{19}) coincides with the speed of light,  $c_s\rightarrow 1$. In this context, we observe that this speed introduces a new constraint on the Rastall parameter, since the speed of sound $c_s\leq 1$. Thus,  we determine that a lower bound for the Rastall parameter is  given by
\begin{equation}
\lambda_\text{Ras}\geq\,1,
\end{equation}
and this suggests that the singular values of the Rastall parameter, $\lambda_{\mathrm{Ras}} \neq 2/3$ and $\lambda_{\mathrm{Ras}} \neq 1/2$, lie outside the validity range of $\lambda_{\mathrm{Ras}}$.

On the other hand, within the framework of single-field inflation in the slow-roll regime, quantum fluctuations of the inflaton field give rise to approximately scale-invariant perturbations. The power spectrum of these curvature perturbations can be expressed in terms of the Hubble parameter and the slow-roll parameter $\epsilon_V$, both evaluated at horizon crossing. In this sense, the scalar power spectrum of curvature perturbations, denoted by $A_s$, can be written as \cite{Garriga:1999vw,DeFelice:2013ar}
\begin{equation}
A_s=\frac{3\kappa H^2}{24\pi^2c_s\epsilon_V}\simeq\frac{\kappa^2V}{24(2\lambda_\text{Ras}-1)\pi^2c_s\epsilon_V}    =\frac{\alpha^3}{(2\lambda_\text{Ras}-1)c_s}A_s^\text{(GR)}.\label{As}
\end{equation}

Furthermore, the scale dependence of the scalar power spectrum is given in terms of the scalar spectral index $n_s$, which is defined as
\begin{equation}
    n_s-1=\frac{d\ln A_s}{d\ln k}=2\eta_V-6\epsilon_V=\alpha^{-1}\left(2\eta_V^\text{(GR)}-6\epsilon_V^\text{(GR)}\label{nsRG}\right).
\end{equation}
Additionally, the transverse-traceless tensor perturbations generated during the inflationary scenario give rise to primordial gravitational waves. These tensor perturbations are characterized by their amplitude, commonly denoted as $A_T$, which is proportional to the square of the Hubble parameter.

In this context, we define an important observational quantity known as the tensor-to-scalar ratio, denoted by $r = A_T / A_s$. This ratio can be expressed as a function of the fundamental inflationary parameters. This ratio  quantifies the relative amplitude of tensor perturbations to scalar perturbations and serves as a key observational constraint on the parameters of the inflationary model. It can be written as
\begin{equation}
    r=\frac{A_T}{A_s}=16c_s\epsilon_V=\frac{c_s}{\alpha}\, r^\text{(GR)}.\label{rr}
\end{equation}
As before, in the limit where the parameter $\alpha \rightarrow 1$, the tensor-to-scalar ratio $r$ coincides with the corresponding GR ratio.

\subsection{Reconstruction}

In this subsection, building on the methodology outlined in Ref.\cite{Chiba:2015zpa}, we detail the reconstruction process of the effective potential $V(\phi)$ in terms of the scalar field $\phi$, within a standard Rastall gravity. This reconstruction is guided by treating the scalar spectral index  $n_s=n_s(N)$, with $N$ the number of $e-$folds 
 as an attractor function. Given that the reconstruction involves the function associated to the effective potential  $V(\phi)$, 
 the first step is to express the scalar spectral index  as a function of the number of $e-$folds. To achieve this, Eq.(\ref{nsRG}) must be reformulated in terms of the potential  as functions of the number of $e-$folds 
$N$
 and their derivatives. Thus, by specifying the spectral index $n_s(N)$, we should derive an expression  for  the effective potential in terms of 
number of $e-$folds 
$N$. Subsequently, using Eq.(\ref{deltaN}), we should determine the relation between the number of $e-$folds 
$N$
 and  the scalar field
$\phi$ i.e., $N=N(\phi)$, which then allows us to reconstruct the effective potential $V(\phi)$
from the attractor $n_s$.

We begin by expressing the scalar spectral index, originally defined in Eq.(\ref{nsRG}) as a function of the number of $e-$folds. In this context, the slow-roll parameters involved can be reformulated as functions of the number $N$, taking into account that 
\begin{equation}
    V_\phi=\frac{dV}{d\phi}=\frac{dV}{dN}\frac{dN}{d\phi}=\kappa\frac{V}{V_\phi}\frac{dV}{dN}=\kappa\frac{V}{V_\phi}V_N\,\,\,\,\Rightarrow \,\,\,\, V_\phi^2=(\kappa VV_N),\,\,\,\,\mbox{and then}\,\,\,\,V_N>0.\label{Vf}
\end{equation}
Here we have used Eqs.(\ref{deltaN}) and (\ref{EqsinSR}). In the following, we will consider that the notation $V_N=dV/dN$, $V_{NN}$ corresponds to $V_{NN}=d^2V/dN^2$, etc.

 In the same way, it is possible to show that
\begin{equation}
    V_{\phi\phi}=\frac{d^2V}{d\phi^2}=\frac{dV_\phi}{d\phi}=\kappa\left(\frac{V^2_N+VV_{NN}}{2V_N}\right).
\end{equation}
Using the expressions for the slow-roll parameters defined in  Eqs.(\ref{p1}) and (\ref{p2}),  the scalar spectral index $n_s$, given by Eq.(\ref{nsRG}), can be rewritten in terms of the effective potential 
$V$ and its derivatives with respect to the number of $e-$folds 
$N$, as follows
\begin{equation}
    n_s-1=\frac{d}{dN}\left(\ln\frac{V_N}{V^2}\right)=\left(\ln\frac{V_N}{V^2}\right)_N.\label{ns1}
\end{equation}
This result aligns with the expression derived in Ref.\cite{Chiba:2015zpa} within the framework of GR, specifically concerning the reconstruction of the potential as a function of the number of $e-$folds. 
Thus, from Eq.(\ref{ns1}), we observe that the expression for the scalar spectral index, when rewritten in terms of the effective potential and its derivative with respect to the number of $e-$folds, remains unaltered in the framework of Rastall gravity.

In order to determine the effective potential $V(N)$, we obtain a first integral of Eq.(\ref{ns1}), given by 
\begin{equation}
    \frac{V_N}{V^2}=\exp\left[\int (n_s-1)dN\right].\label{22}
\end{equation}
Now performing a new integration of Eq.(\ref{22}) to obtain  the potential $V(N)$, we find 
\begin{equation}
    V(N)=\left[-\int\left(\exp \left[\int(n_s-1)dN\right]\right)dN\right]^{-1}.\label{VN}
\end{equation}
Thus, in general terms,  Eq.(\ref{VN}) allows us to determine the potential  $V(N)$ given the scalar spectral index  as a function of the number of $e-$folds, i.e., $n_s(N)$.

As we can see from Eq.(\ref{VN}), the reconstruction of the potential in terms of the number of $e-$folds $N$ does not depend on Rastall gravity and coincides with the result obtained in GR \cite{Chiba:2015zpa}.

However, as we will show later, the reconstruction of the effective potential as a function of the scalar field is indeed affected by Rastall theory. Moreover, the parameter space is also constrained by the theory, due to observational bounds on scalar perturbations and the tensor-to-scalar ratio, as given by Eqs.(\ref{As}) and 
(\ref{rr}).

\subsection{An Example}

In this subsection, we implement the above formalism in the context of our reconstruction in the Rastall inflation scenario with constant $\lambda_\text{Ras}=$ , using the simplest form of the scalar spectral index as a function of the number of $e-$ times $N$
 to analytically reconstruct the effective potential
$V(\phi)$. In this sense, we assume that the scalar spectral index $n_s(N)$ is given by the attractor \cite{Starobinsky:1980te,Chiba:2015zpa,Kallosh:2013hoa}
\begin{equation}
n_s(N)=1-\frac{2}{N}.\label{ns}
\end{equation}
Specifically,  assuming that the horizon exit occurs approximately at 
$N=60$  before the end of inflation, the resulting scalar spectral index (attractor) proposed from the relation (\ref{ns}) is found to be in good agreement with current observational data.

In this form, combining  Eq.(\ref{VN}) with (\ref{ns}), we find that the effective potential in terms of the number of $e-$folds $N$ becomes
\begin{equation}
V(N)=\left[\frac{A}{N}+B\right]^{-1},\label{V1}
\end{equation}
where $A$ and $B$ are two integration constants. Here, we note that the constant $A$ is positive, as indicated by the expression $V_N=A\,V^2/N^2>0$. This condition arises from the relationship $V_\phi^2=(\kappa VV_N)$, and given that $V>0$ and $\kappa>0$, it follows that $V_N=V_\phi^2/(\kappa V)>0$ (refer to Eq.(\ref{Vf})), while $B$ is an integration constant with an arbitrary sign.

 Now from Eqs.(\ref{As}), (\ref{rr}) and (\ref{V1}), we find that the constraints on the integration constants $A$ and $B$ are given by
\begin{equation}
A=\frac{\kappa^2\,N_*^2}{12\,\pi^2\,(2\lambda_\text{Ras}-1)\,c_s A_{s*}},\,\,\,\,\,\,\,\mbox{and}\,\,\,\,\,\,\,\,\,B=\frac{\kappa^2\,(8-(r_*/c_s)\,N_*)}{12\,\pi^2\,(2\lambda_\text{Ras}-1)\,A_{s*}\,r_*},
\end{equation}
respectively.
Here, the number $N_*$ denotes the number of $e-$folds at the epoch when the cosmological
scale exits the horizon and  the quantities $A_{s*}$ and $r_*$ correspond to the observational values of the scalar power spectrum and the tensor-to-scalar ratio, respectively, evaluated at this epoch. In particular, considering that at $N_*=60$ the scalar power spectrum  is $A_{s*}=2.2\times 10^{-9}$ and the upper limit for the tensor-to-scalar ratio  $r_{*}=0.039$, we find the constraints
\begin{equation}
A\simeq 8.7\times 10^{12}\,\frac{(2\lambda_\text{Ras}-1)}{\sqrt{\lambda_\text{Ras}}(3\lambda_\text{Ras}-2)^{3/2}\,M_p^4},\,\,\,\,\mbox{and}\,\,\,\,B\simeq 5.0\times 10^{11}\left(1-\frac{0.3}{\sqrt{\lambda_\text{Ras}}}\sqrt{3\lambda_\text{Ras}-2}\right)\,\frac{(2\lambda_\text{Ras}-1)}{(3\lambda_\text{Ras}-2)^2\,M_p^4},\label{AB}
\end{equation}
respectively. In the specific case in which the parameter $\lambda_\text{Ras}=1$, the constraints on 
$A$ and 
$B$ coincide with those obtained within the framework of  GR. 

On the other hand, to reconstruct the effective potential as a function of the scalar field, we first consider the relation between the scalar field and the number of $e$-folds given by Eq.(\ref{deltaN}), in the case where the integration constant $B > 0$, results
\begin{equation}
N=\frac{A}{B}\,\sinh^2\left[\sqrt{\frac{2\pi B}{A}\left(\frac{2\lambda_\text{Ras}-1}{3\lambda_\text{Ras}-2}\right)}\frac{(\phi-C)}{M_p}\right].
\end{equation}
 In this way,  the reconstructed effective potential 
 $V(\phi)$ in terms of the scalar field can be expressed as 

\begin{equation}
V(\phi)=V_0\,\tanh^2\left[\sqrt{\frac{2\pi B}{A}\left(\frac{2\lambda_\text{Ras}-1}{3\lambda_\text{Ras}-2}\right)}\frac{(\phi-C)}{M_p}\right],\,\,\,\,\,\mbox{with}\,\,\,\,\,\,V_0=1/B,
\label{POR}
\end{equation}
and the quantity $C$ denotes a new integration constant. Here, we note that the effective potential as a function of the scalar field depends on the Rastall parameter. In the special case in which the Rastall parameter $\lambda_\text{Ras}=1$, the effective potential given by Eq.(\ref{POR})  reduces to that of  GR, see Refs.\cite{Chiba:2015zpa,Kallosh:2013hoa}.

\begin{figure}[ht]
    \centering
\includegraphics[width=0.85\linewidth]{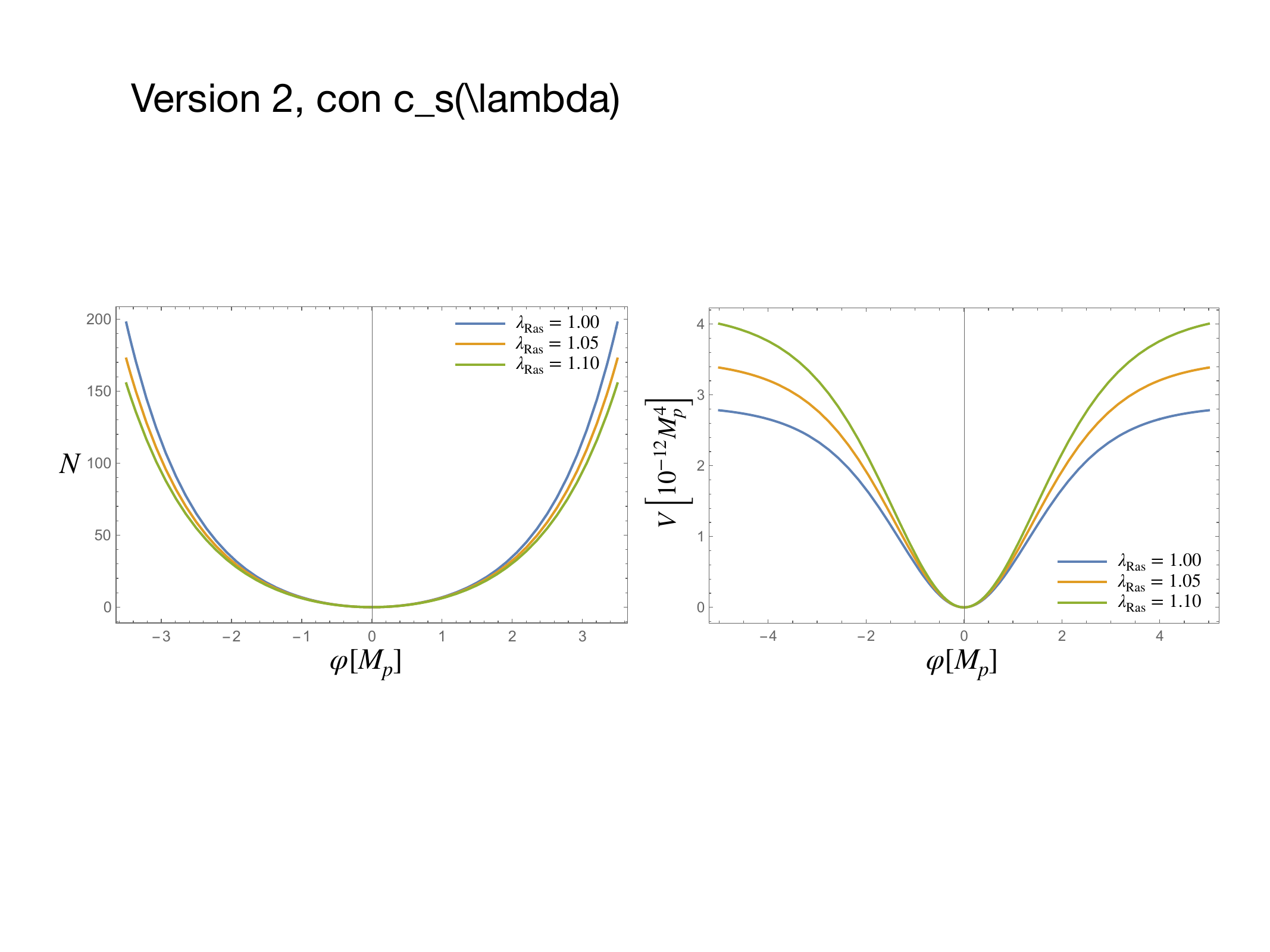}
    \caption{The left panel illustrates the number of $e$-folds $N$ as a function of the shifted scalar field $\varphi=\phi- C$. The right panel shows the reconstructed effective potential as a function of the scalar field $\varphi$. In both cases, we have considered three distinct values of the Rastall parameter $\lambda_\text{Ras}$; 1.00 (blue curve), 1.05 (orange curve), and 1.10 (green curve).} 
    \label{Potencial1}
\end{figure}

In Fig.\ref{Potencial1}  the left panel shows the number of $e$-folds $N$  as a function of the shifted field $\varphi =\phi-C$. The right panel shows the reconstructed effective potential as a function of $\varphi$. In both panels, we consider three different values of the Rastall parameter $\lambda_\text{Ras}$: 1.00 (blue curve), 1.05 (orange curve), and 1.10 (green curve), respectively. In addition, we have used the values of the parameters $A$ and $B$ given by Eq.(\ref{AB}).

From the figure, we can see that the inflationary scenario could take place through two possibilities: starting from positive values of the field $\varphi$ or from negative ones. In both cases, the inflationary epoch ends when the field approaches values close to zero. We also note that for large values of $|\varphi|$ ($|\varphi| > 5M_p$), the effective potential exhibits a flat region where the number of $e-$folds  exceeds  300 ($N>300$) for values of $\lambda_\text{Ras}\approx 1$. In addition, from the figure we observe that for the value of the Rastall parameter $\lambda_{\text{Ras}} = 1$ (GR limit), the number of $e$-folds as a function of the scalar field is higher compared to the case when $\lambda_{\text{Ras}} > 1$.

\section{Generalized Rastall gravity}\label{Gen}

Let us now examine the framework of generalized Rastall gravity, in which the Rastall parameter $\lambda_\text{Ras}$, originally a constant, is generalized to be a function of the space-time coordinates, that is, $\lambda_\text{Ras}=\lambda_\text{Ras}(x^\mu)$. In particular, we may assume that the variation of the parameter $\lambda_{\text{Ras}}$ is homogeneous on large scales and can be treated as a function of time (or the scalar field), such that $\lambda_{\text{Ras}}=\lambda_{\text{Ras}}(\phi(t))$.
In this context, we  can generalize  Eq.(\ref{noncT}) associated to the covariant divergence of the energy-momentum tensor as \cite{Moradpour:2017shy,Das:2018dzp}
\begin{eqnarray}
\nabla_\mu T^{\mu\nu}= \nabla^\nu\left[\frac{(1-\lambda_\text{Ras})}{2\kappa} R\right]\propto \nabla^\nu[ (1-\lambda_\text{Ras})R/\kappa],\label{T1}
\end{eqnarray}
and this modification corresponds to the generalized Rastall theory of gravity. Here, we note that,  the gravitational field equation given in Eq.(\ref{campoRas}) remains unchanged, but the Rastall parameter becomes variable.

Furthermore, within this framework, Eqs.(\ref{campoRas}) and (\ref{Teff}) are modified by considering that the term $\kappa(\lambda_{\text{Ras}})$, given by Eq.(\ref{kras}), is not a constant. 
In this context, from the energy-momentum tensor defined in Eq.(\ref{Teff}), we can identify the energy density and pressure associated with a non-canonical scalar field 
 within the framework of the generalized Rastall theory, where the Rastall parameter
$\lambda_{\text{Ras}}=\lambda_{\text{Ras}}(\phi)$ is not a constant and $\rho$ and $p$ can be written as 
\begin{eqnarray}
    \rho=\frac{K_1(\phi)}{2}\dot{\phi}^2+\tilde{V}(\phi),\,\,\,\,\,\,\,\;\;\mbox{and}\;\;\;\;\;\;\;\,
    p=\frac{K_2(\phi)}{2}\dot{\phi}^2-\tilde{V}(\phi),
\end{eqnarray}
where the functions $K_1(\phi)$,  $K_2(\phi)$ and the redefined effective potential $\tilde{V}(\phi)$ are given by
\begin{equation}
K_1(\phi)=\frac{3\lambda_\text{Ras}(\phi)-2}{2\lambda_\text{Ras}(\phi)-1},\,\,\,\,  K_2(\phi)=\frac{\lambda_\text{Ras}(\phi)}{2\lambda_\text{Ras}(\phi)-1},\,\,\,\, \mbox{and}\,\,\,\, \tilde{V}(\phi)=\frac{V(\phi)}{2\lambda_\text{Ras}(\phi)-1},\label{K2}
\end{equation}
respectively. 

In relation to the effective speed of sound, it is variable and defined as  \cite{Garriga:1999vw,DeFelice:2013ar,Herrera:2024ojo}
\begin{equation}
    c_s^2(\phi)=\frac{p_X}{\rho_X}=\frac{\lambda_\text{Ras}(\phi)}{3\lambda_\text{Ras}(\phi)-2}.\label{cs2}
\end{equation}
As before, in the special case where the variable parameter $\lambda_\text{Ras}\rightarrow 1$, the speed $c_s$ reduces to the speed light, $c_s\rightarrow 1$. 

In addition,  from Eq.(\ref{cs2}) it is important to observe that the square of the velocity of sound $c_s^2\leq 1$ and consequently the Rastall parameter obeys the lower bound $\lambda_\text{Ras}\geq1$.

On the other hand,  the curvature perturbation power spectrum can be expressed in terms of the Hubble, the parameter $\kappa$ associated to the Rastall parameter, the speed $c_s$ and slow-roll parameters, all evaluated at horizon crossing. In this context, the scalar power spectrum  $A_s$, is defined as \cite{Garriga:1999vw,DeFelice:2013ar,Armendariz-Picon:1999hyi,Herrera:2024ojo}
\begin{equation}
A_s=\frac{3\kappa H^2}{24\pi^2c_s\epsilon_V},\label{As2}
\end{equation}
where the slow-roll parameter $\epsilon_V$ in this framework is defined as
\begin{equation}
 \label{eLnoc}   \epsilon_V=\frac{\tilde{f}(\phi)}{2\kappa(\phi)}\left[\tilde{f}(\phi)-\frac{\lambda_{\text{Ras}_\phi}}{(3\lambda_\text{Ras}(\phi)-2)(2\lambda_\text{Ras}(\phi)-1)}\right],\,\,\,\,\mbox{with}\,\,\,\,\tilde{f}(\phi)=\frac{V_\phi}{V}-\frac{2\lambda_{\text{Ras}\,\phi}}{\left(2\lambda_\text{Ras}(\phi)-1\right)},
\end{equation}
and $\lambda_{\text{Ras}\,\phi}=\partial\lambda_\text{Ras}/\partial\phi$. Here, we note that when  $\lambda_\text{Ras}(\phi)=$constant =1, the slow roll parameter $\epsilon_V$ reduces to the standard slow roll parameter in GR $\epsilon_V=(V_\phi/V)^2/(2\kappa^\text{(GR)})$.

 The scalar spectral index $n_s$ can be determined from the definition $n_s-1=\frac{d A_s}{d \ln k}$, with $A_s$  defined by Eq.(\ref{As2}). Thus, we find that 
this observational parameter can be written as
\begin{equation}
    n_s-1=\frac{d A_s}{d \ln k}\simeq -(2\epsilon+\eta+s+\delta),\label{nsm12}
\end{equation}
where the slow-roll parameters $\epsilon$, $\eta$, $s$ and  $\delta$  are given by
\begin{equation}
    \epsilon=-\frac{\dot{H}}{H^2},\,\,\,\,\,\,\eta=\frac{\dot{\epsilon}}{H\epsilon},\,\,\,\,\,\, s=\frac{\dot{c}_s}{Hc_s}\,\,\,\,\,\;\mbox{and}\,\,\,\, \;\;\delta=\frac{\dot{\kappa}}{H\kappa},
\end{equation}
respectively. Here, we introduce the slow-roll parameters $s$ and $\delta$, which characterize the variation of the speed of sound $c_s$ and the Rastall parameter $\lambda_{\text{Ras}}$, respectively.

The tensor-to-scalar ratio is defined as \cite{Garriga:1999vw,DeFelice:2013ar}
\begin{equation}
    r=\frac{A_T}{A_s}=16c_s\epsilon_V.\label{rr2}
\end{equation}
As before, this ratio characterizes the relative contribution of tensor and scalar perturbations and provides a central observational constraint on the inflationary parameters of the model.

On the other hand, to calculate the scalar spectral index and these parameters, we can rewrite the slow-roll parameter defined by Eq.(\ref{eLnoc}) as follows:
\begin{equation}
    \epsilon_V=\frac{1}{2\kappa}\left[u^2+fu\right],
\end{equation}
in which the functions $u(\phi)$ and $f(\phi)$ are given by 
\begin{equation}
    u=u(\phi)=\frac{\tilde{V}_\phi}{\tilde{V}},\,\,\,\,\,\,\,\,\,\,\,\,\mbox{and}\,\,\,\, \,\,\,\,\,\,\,\,\,f=f(\phi)=\frac{d\ln \kappa}{d\phi},
\end{equation}
respectively. In addition, the slow-roll parameters $\eta$, $s$, and $\delta$ can be rewritten as follows.
\begin{equation}
    \eta=\frac{1}{\kappa}\left[u\left(f-\frac{d\ln (u+f)}{d\phi}\right)-u_\phi\right],\,\,\,\,\,\,\, s=-\frac{f\,u(2\lambda_\text{Ras}-1)}{\kappa\lambda_\text{Ras}},\,\,\,\,\,\,\mbox{and}\,\,\,\,\,\,\,\,\delta=-\frac{u}{\kappa}f,
\end{equation}
respectively.

In this form, we find that the scalar spectral index defined by Eq.(\ref{nsm12}) can be written as 
\begin{equation}
n_s\simeq1-\frac{d}{dN}\ln\left[\frac{\kappa^3\lambda_\text{Ras}\tilde{V}}{u(u+f)}\right].
\end{equation}

Thus, the first integral of the above equation becomes
\begin{equation}
\left[\frac{\kappa^3\lambda_\text{Ras}\tilde{V}}{u(u+f)}\right]\simeq\exp\left[\,\int(1-n_s)dN\,\right].\label{ec47}
\end{equation}

In order to rewrite this equation, we consider the relation between the number of $e-$folds and the scalar field, given by $dN=-Hdt=-(H/\dot{\phi})d\phi$. 
Here, we note that this relation does not depend on the gravitational theory or the specific inflationary model, because it is derived only from the definition of the number of $e$-folds and the definition of the Hubble parameter. In this sense, the number of $e$-folds measures how much the scale factor grows during inflation $N-N_{end}=\ln[a_{end}/a(t)]$ (see Eq.(\ref{deltaN})) which is purely kinematical and does not field equations or gravitational theory assumptions are utilized. Thus, from this relation we can written 
\begin{equation}
dN=\frac{\kappa}{u}d\phi,\qquad f=\frac{d\ln \kappa}{d\phi}=\frac{\kappa}{u}\frac{d\ln \kappa}{dN},\,\,\,\,\,\,\,\,\,\mbox{with}\qquad u^2=\kappa\,\frac{\tilde{V}_N}{\tilde{V}},\,\,\,\,\,\,\,\,\mbox{thereby}\qquad \tilde{V}_N>0.\label{dNd}
\end{equation}

In this way, from Eq.(\ref{ec47}) the differential equation for the effective potential as a function of the number of $e$-folds can be reconstructed from the parametrizations of the scalar index $n_s(N)$ and the Rastall parameter $\lambda_\text{Ras}(N)$ as
\begin{equation}
\frac{\tilde{V}_N}{\tilde{V}}+\frac{d\ln \kappa}{dN}=\kappa^2\lambda_\text{Ras}\tilde{V}\,\exp\left[\int (n_s-1)dN\right].\label{dif}
\end{equation}
It is worth highlighting that, when $\lambda_\text{Ras}=1$ is taken as a constant, Eq.(\ref{dif}) recovers the GR limit. Thus, Eqs.(\ref{dNd}) and (\ref{dif}) are fundamental to reconstruct the potential as a function of the scalar field, for a given parameterization of the spectral index and the Rastall parameter in terms of the number of $e-$ times $N$. In the following, we provide an example of a parametrization of the spectral index and the Rastall parameter, in order to reconstruct the effective potential in terms of the scalar field.

\subsection{An example}
In this subsection, we apply the above formalism to our reconstruction within the framework of generalized Rastall inflation, assuming the simplest dependence of the Rastall parameter $\lambda_\text{Ras}$ on the number of $e$-folds $N$ together with the attractor given by Eq.(\ref{ns}) for the scalar spectral index $n_s=n_s(N)$. In this sense, we provide an analytical reconstruction of the effective potential $V(\phi)$.

In this context, we can consider that the Rastall parameter $\lambda_\text{Ras}$ evolves slowly  with the number of $e$-folds according to 
\begin{equation}
\lambda_\text{Ras}(N)=1+\beta\,N, \,\,\,\,\,\,\mbox{with the dimensionaless parameter }\,\,\,\,\beta\ll 1.\label{ec2}
\end{equation}
In particular, we consider that at the end of inflation, when the number of $e$-folds satisfies $N_{\text{end}} \sim 0$, the Rastall parameter approaches $\lambda_\text{Ras} \simeq 1$, which coincides with the GR limit. Regarding the parametrization of the Rastall parameter given by Eq.(\ref{ec2}) in terms of the number of $e$-folds $N$, we note that there is no study addressing the reconstruction of an inflationary stage based on this parametrization in the literature. In this form, the motivation for considering the simplest parametrization of the Rastall parameter, described by Eq.(\ref{ec2}), is to account for small deviations from GR (corresponding to $\lambda_\text{Ras} = 1$), controlled by the parameter $\beta$ during inflation, where the number of $e-$folds $N$ varies approximately between $0\lesssim N\lesssim \mathcal{O}(10^2) $. Besides, observational data, particularly from cosmology and local gravity tests, constrain the Rastall parameter $\lambda_\text{Ras}$ to be very close to the unity, ensuring consistency with GR \cite{Bishi:2023mwv,Tang:2019dsk,daSilva:2020okh}. In addition, other  analyzes based on CMB, BAO, galaxy rotation curves, gravitational lensing, and compact objects consistently show that only small deviations from the GR limit are allowed, see also  \cite{Akarsu:2020yqa,Oliveira:2015lka}.
 In most formulations of Rastall gravity, viable cosmological and astrophysical scenarios require this parameter to lie very close to the value that recovers GR, i.e., $\lambda_\text{Ras}\sim 1$. This suggests that regimes in which the Rastall parameter takes very large values ($\lambda_\text{Ras}\gg 1$) even under the assumption of large  number of $e-$folds $N$, are strongly disfavored by current gravitational tests.

In order to reconstruct the effective potential, we consider the parameter $n_s(N)$ from Eq.(\ref{ns}) and the Rastall parameter $\lambda_\text{Ras}(N)$ (or $\kappa(N)$) from Eq.(\ref{ec2}), both expressed as functions of the number of $e$-folds. In this form, from Eq.(\ref{dif}), we obtain the first-order differential equation governing the effective potential, which takes the form

\begin{equation}
\left[\frac{\kappa^3\lambda_\text{Ras}\tilde{V}}{u(u+f)}\right]=AN^2,\,\,\,\,\,\,\Rightarrow\,\,\,\,\frac{\tilde{V}_N}{\tilde{V}}=-(\ln \kappa)_N+\left[\frac{\kappa^2\lambda_\text{Ras}\tilde{V}}{AN^2}\right],\,\,\,\,\mbox{with}\,\,\,(\ln \kappa)_N=-\frac{(\lambda_\text{Ras})_N}{(2\lambda_\text{Ras}-1)(3\lambda_\text{Ras}-2)}\label{ERR},
\end{equation}
and $A$ a dimensionless integration constant. We note that, in the special case where $\lambda_\text{Ras}=1$ (constant), Eq.(\ref{ERR}) reduces to the GR limit in which $V_N/V^2\propto N^{-2}$.

Thus, from Eq.(\ref{ERR}), we find that the solution for the effective potential as a function of the number of $e$-folds $N$ takes the form of
\begin{equation}
\tilde{V}(N)=\frac{N(1+3\beta N)}{(1+2\beta N)\left[\kappa^\text{(GR)2}A^{-1}+NB-\frac{2\beta N}{3A}\kappa^\text{(GR)2}\ln(1+3\beta N)\right]},
\end{equation}
where $B$ corresponds to a new integration constant with dimension $M_p^{-4}$. Thus, the original effective potential $V$ in terms of the number of $e-$folds can be written as
\begin{equation}
V(N)=\frac{N(1+3\beta N)}{\left[\kappa^\text{(GR)2}A^{-1}+NB-\frac{2\beta N}{3A}\kappa^\text{(GR)2}\ln(1+3\beta N)\right]}.\label{VN2}
\end{equation}
In addition, we note that in the limit $\beta = 0$, that is, when $\lambda_\text{Ras}$ is constant and equal to 1, we recover the expression for the potential obtained in GR, see Ref.\cite{Chiba:2015zpa}. 
Now, in order to reconstruct the effective potential as a function of the scalar field, we require to establish the relation between the number of $e$-folds and the scalar field, i.e. $N = N(\phi)$. In this context,  to determine the relation between the number of $e-$folds and the field $\phi$, we can write the first term of  Eq.(\ref{dNd}) as    
\begin{equation}
\sqrt{\frac{\tilde{V}_N}{\kappa\,\tilde{V} }}\,\,dN= \,d\phi.\label{ddd}
\end{equation}
Thus, we need to substitute the potential expressed in terms of the number of $e$-folds, as defined in Eq.(\ref{VN2}), together with the Rastall parameter given in Eq.(\ref{ec2}), into Eq.(\ref{ddd}). However, it is not possible to obtain an analytical solution for $N = N(\phi)$. Thus, in order to obtain an analytical solution to Eq.(\ref{ddd}) and then $N=N(\phi)$, we assume that the parameter $\beta$ associated with the Rastall parameter is small, allowing us to expand the root of the equation.
In this context, assuming that the parameter $\beta \ll 1$ and $B>0$, Eq.(\ref{ddd}) can be approximated as 
\begin{equation}
\left\{\frac{1}{\sqrt{AN}}\left(\frac{\kappa^\text{(GR)}}{A}+
\frac{B\,N}{\kappa^\text{(GR)}}\right)^{-1/2}\,\left[1+\,\left(1+
\frac{A\,B\,N}{2\kappa^\text{(GR)2}}\right)\beta N\,+\,\cdots\right]     \right\}\,dN=d\phi.
\end{equation}

By solving this differential equation, we find that the number of $e$-folds as a function of the field $\phi$ can be written as 
\begin{equation}
    N(\phi)= \mathcal{F}^{-1}(\phi),\label{N2}
\end{equation}
where $\mathcal{F}^{-1}(\phi)$ denotes the inverse function of 
$$
\mathcal{F}(\phi)=\frac{\sqrt{\kappa^\text{(GR)}}}{4\,A\,B^{3/2}}\,\Bigg[\sqrt{A}\,\left(16B-5\kappa^\text{(GR)2}A^{-1}\,\beta\right)\,\mbox{arcsinh}\left(\frac{\sqrt{AB\,d(\phi)}}{\kappa^\text{(GR)}}\right)
$$
\begin{equation}
+\,\,\left(\beta\sqrt{B\,d(\phi)}A/\kappa^\text{(GR)}\right)\left(5\kappa^\text{(GR)2}A^{-1}+2B\,d(\phi)\right)\,\sqrt{1+\frac{AB\,d(\phi)}{\kappa^\text{(GR)2}}}\,\Bigg],\label{sl2}
\end{equation}
where the dimensionless function $d(\phi)$ associated to scalar field  is defined as 
\begin{equation}
d(\phi)=C+2\sqrt{\kappa^\text{(GR)}}\,\phi,\label{d2}
\end{equation}
and $C$ a new integration constant. Here we note that in the GR limit with $\beta = 0$ (or $\lambda_\text{Ras} = 1$), Eq.(\ref{sl2}) reduces to the standard GR result,
$N(\phi)\propto\mathcal{F}^{-1}(\mbox{arcsinh}[(AB\,d(\phi)/\kappa^\text{(GR)2})^{1/2}])\propto\mbox{sinh}^{2}[(AB\,d(\phi)/\kappa^\text{(GR)2})])$ \cite{Chiba:2015zpa}.

In this way, by substituting the number of $e$-folds expressed in terms of the scalar field from Eq.(\ref{sl2}) into the potential of Eq.(\ref{VN2}) under the approximation $\beta\ll 1$, we obtain the reconstructed effective potential as a function of the scalar field becomes
\begin{equation}
V(\phi)\approx\,\frac{\mathcal{F}^{-1}(\phi)}{\kappa^\text{(GR)\,2}A^{-1}+B\mathcal{F}^{-1}(\phi)}\left[1+\beta\,\mathcal{F}^{-1}(\phi)\right].\label{VV2}
\end{equation}
Here, as before, we note that in the limit $\beta = 0$, the first term of the reconstructed potential $V(\phi)$ coincides with the effective potential obtained within the GR framework.

In addition, we find that the Rastall parameter in terms of the scalar field  becomes
\begin{equation}
\lambda_\text{Ras}(\phi)=1+\beta\,\mathcal{F}^{-1}(\phi),\label{LL2}
\end{equation}
where we have utilized Eqs.(\ref{ec2}) and (\ref{N2}), respectively.

In the other hand, in order to determine the constraints on the different parameters in our generalized Rastall gravity, we make use of the observational cosmological parameters. In this sense, from Eqs.(\ref{As2}) and (\ref{rr2}), we find that the scalar power spectrum and the tensor-to-scalar ratio in terms of the number of $e-$folds $N$ become
\begin{equation}
    A_s(N)=\frac{\kappa^2 \tilde{V}^2}{12\pi^2}\left(\frac{1+3\beta N}{1+\beta N}\right)^{1/2}\left(\tilde{V}_N+\tilde{V}\frac{d\ln\kappa}{dN}\right)^{-1}=\frac{A\,N^2}{12\pi^2}\left[\frac{(1+3\beta\,N)}{(1+\beta\,N)^3}\right]^{1/2},\label{As3} 
\end{equation}
and for the tensor-to-scalar ratio as a function of number $e-$folds we get
\begin{equation}
    r(N)=8\left(\frac{1+\beta N}{1+3\beta N}\right)^{1/2}\left(\frac{\tilde{V}_N}{\tilde{V}}+\frac{d\ln\kappa}{dN}\right)=\frac{24(1+2\beta N)}{N[3(1+ABN/\kappa^\text{(GR)\,2})-2\beta N\,\ln[1+3\beta N]]}\left[\frac{(1+\beta N)}{(1+3\beta N)}\right]^{3/2}.\label{r22} 
\end{equation}

It can be observed that when $\beta=0$ ($\lambda_\text{Ras}=1$), the conventional expressions of $A_s(N)$ and $r(N)$ are obtained within the framework of general relativity theory; see Ref.\cite{Chiba:2015zpa}.

From Eqs.(\ref{As3}) and (\ref{r22}), we find that the constraints on the parameters $A$ and $B$, are given by
\begin{equation}
A=\frac{12\pi^2 A_{s*}}{N_*^2}\left[\frac{(1+\beta\,N_*)^3}{(1+3\beta\,N_*)}\right]^{1/2}, \label{A2}
\end{equation}
and 
\begin{equation}
   B=\frac{\kappa^\text{(GR)\,2}}{18\pi^2 r_* A_{s*}(1+3\beta N_*)}\left\{12(1+2\beta N_*)+\left[\beta r_* N^2_{*} \ln(1+3\beta N_*)-\frac{3}{2}r_*N_*\right]\left[\frac{(1+3\beta\,N_*)}{(1+\beta\,N_*)}\right]^{3/2}\right\}, \label{B2}
\end{equation}
respectively.
As before, the number $N_*$ corresponds to the number of $e-$folds at the epoch when the cosmological
scale exits the horizon and  the values  $A_{s*}$ and $r_*$ denote the observational values of the scalar power spectrum and the tensor-to-scalar ratio, respectively, evaluated at this stage. In particular, considering that at $N_*=60$ the scalar power spectrum is $A_{s*}=2.2\times 10^{-9}$ and the upper limit for the tensor-to-scalar ratio $r_{*}=0.039$, we obtain for the specific value of $\beta=10^{-2}$, the constraints on the integration constants $A$ and $B$ given by $A=8.75\times10^{-11}$ and $B=5.07\times10^{8}\kappa^\text{(GR)\,2}$. 
For $\beta=10^{-3}$, we obtain the values; $A=7.27\times10^{-11}$ and $B=5.20\times10^{8}\kappa^\text{(GR)\,2}$. For the case in which $\beta=10^{-4}$, we find that the parameters $A$ and $B$ are $A=7.24\times10^{-11}$ and $B=5.52\times10^{8}\kappa^\text{(GR)\,2}$, respectively. For these values of the parameter $\beta$, we note that the constraints on the integration constants $A$ and $B$ are of the order of $A\sim 10^{-10}$ and $B\sim  10^{9}\,\kappa^\text{(GR)\,2}$, respectively.We also note that for these different values of $\beta$, the constraints obtained on the parameters $A$ and $B$ are similar.
Thus, the nearly identical constraints obtained for the integration constants $A$ and $B$ across different values of $\beta$ indicate a slight  degeneracy among the parameters. This suggests that the current data are insensitive to the individual contributions of 
$A$ and 
$B$, even when allowing for variations in the parameter 
$\beta$. Moreover, this outcome is model-dependent, since we have assumed  a parametrization on the Rastall parameter in which
 $\lambda_\text{Ras}\sim 1$ when $\beta\ll1$, see Eq.(\ref{ec2}).

\begin{figure}[ht]
    \centering
\includegraphics[width=0.5\linewidth]{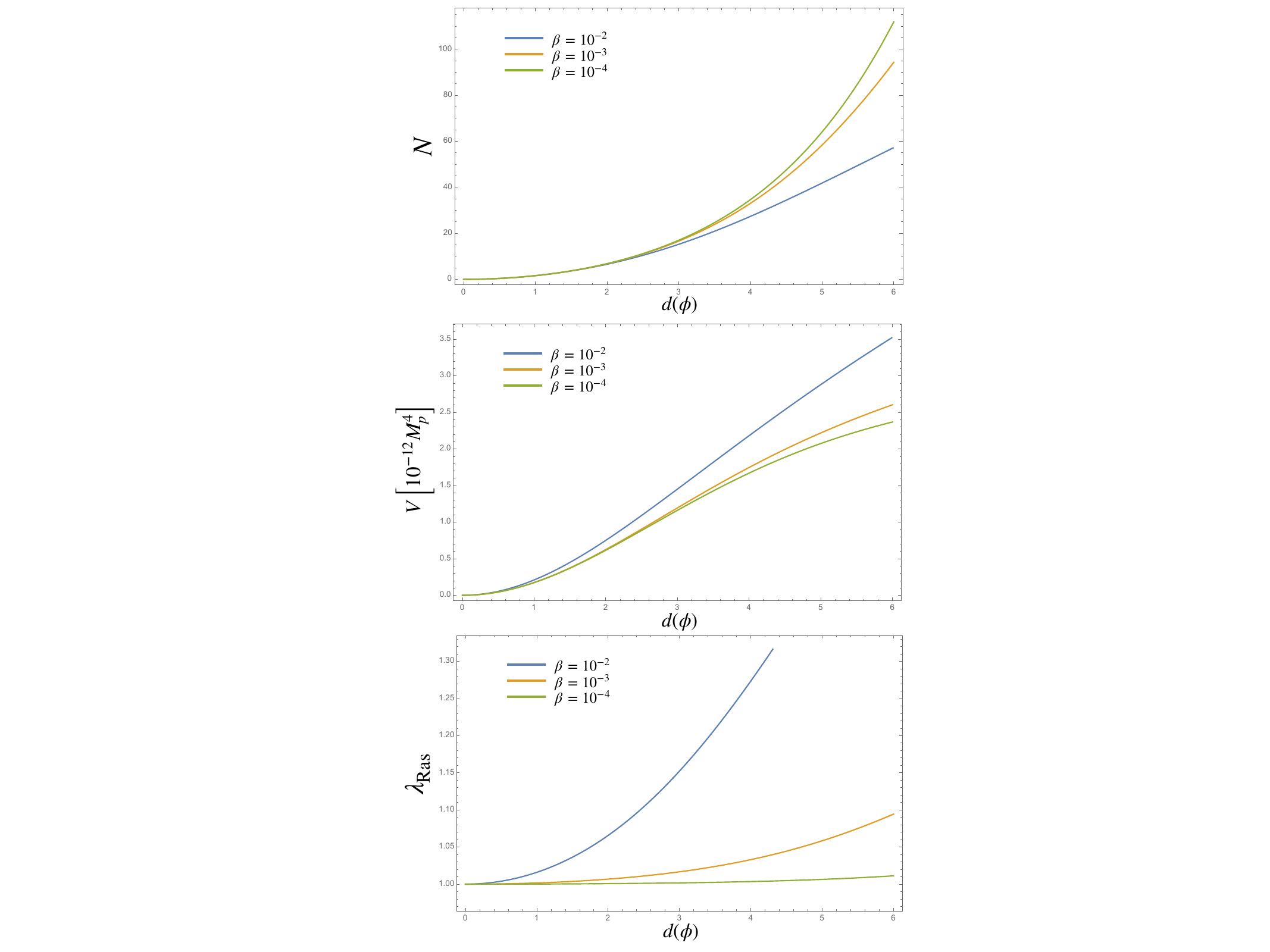}
    \caption{Dependence of the number of $e$-folds (upper panel), the reconstructed effective potential (central panel), and the Rastall parameter (lower panel) on the scalar field during the inflationary epoch. In these plots, we have considered three distinct values of the parameter $\beta$, together with the corresponding constraints on the parameters $A$ and $B$, as provided by Eqs.(\ref{A2}) and (\ref{B2}), respectively.} 
    \label{fig2}
\label{Potencial2}
\end{figure}

In Fig.\ref{fig2}, we present the evolution of the number of $e$-folds $N(\phi)$ (upper panel), the reconstruction of the effective potential $V(\phi)$ (central panel), and the Rastall parameter $\lambda_\text{Ras}(\phi)$ as a function of the scalar field, in terms of the dimensionless function $d(\phi)$ defined in Eq.(\ref{d2}). In all panels, we have considered three different values of the parameter $\beta$, together with the corresponding constraints on the parameters $A$ and $B$ given by Eqs.(\ref{A2}) and (\ref{B2}), respectively. In order to write down the number of $e-$folds as a function of the scalar field, we utilize Eq.(\ref{N2}), where the function $\mathcal{F}(\phi)$ is defined by Eq.(\ref{sl2}). From this panel, we observe that the number of $e$-folds $N(\phi)$ increases as $\beta$ takes progressively smaller values, eventually approaching the GR limit characterized by $\lambda_\text{Ras}\simeq 1$.
The central panel shows the evolution of the reconstructed effective potential $V(\phi)$, obtained from Eq.(\ref{VV2}). Here, we note that during the inflationary scenario the reconstructed effective potential takes the value $V \sim 10^{-12}M_p^4$, and that it increases as $\beta$ becomes larger. The lower panel shows the evolution of the Rastall parameter in terms of the scalar field given by Eq.(\ref{LL2}). Here, we note that at the end of the inflationary stage, when the number of $e$-folds $N\simeq 0$, the Rastall parameter reduces to that of GR gravity. Also, we observe that for the values $\beta=10^{-4}$ and $\beta=10^{-3}$, the Rastall parameter does not exhibit significant evolution and remains practically constant during the inflationary epoch.

\begin{figure}[ht]
    \centering
\includegraphics[width=0.5\linewidth]{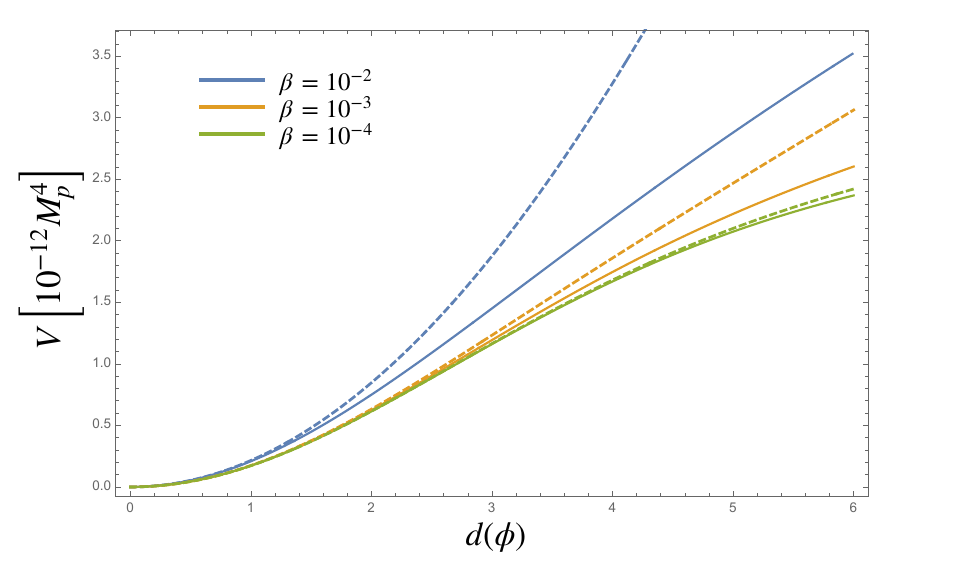}
    \caption{The plot illustrates the comparison between the numerical solution of Eq.(\ref{ddd}) together with Eq.(\ref{VN2})(dashed line) and the approximate analytical solution of the reconstructed potential $V(\phi)$ provided by Eq.(\ref{VV2}) under the condition $\beta\ll 1$ (solid line). Here we have considered three different values of the parameter $\beta$.}
    
\label{CompNum}
\end{figure}

In Fig.\ref{CompNum}, we show the comparison between the effective potentials $V(\phi)$, considering  the numerical solution of Eq.(\ref{ddd}) together with Eq.(\ref{VN2})  (dashed line) and the approximate analytical expression for the reconstructed potential $V(\phi)$ given by Eq.(\ref{VV2}), derived under the assumption $\beta \ll 1$ (solid line). As in Fig.\ref{fig2}, we have considered  the three values for the parameter  $\beta$. This figure  shows that the analytical approach in which $\beta=10^{-4}$, both potentials (numerical and approximate) are very similar (green line) during full inflationary scenario.   For the situation $\beta=10^{-3}$, we note that  the approximation for the reconstructed potential is not valid for values of the number of $e-$folds $N\gtrsim$ 40.
Quantitatively, the most significant deviations arise for $\beta \simeq \mathcal{O}(10^{-2})$, where the small $\beta$ expansion used to obtain Eq.(\ref{VV2}) begins to lose accuracy. Since the analytical potential neglects higher order contributions in $\beta$, its precision naturally deteriorates as $\beta$ approaches the upper edge of the regime $\beta \ll 1$. Since the reconstruction of the effective potential $V(\phi)$ is based on the condition $\beta\ll 1$, the analytical accuracy of the reconstructed potential breaks down as the parameter $\beta$ increases.
In this sense,  Fig.\ref{CompNum}  shows that 
the numerical results validate the analytical approximation within its intended domain of applicability.

\section{Stability analysis}\label{SA}
In this section, we study the stability analysis for our reconstructed models in the framework of standard and generalized Rastall gravity. To examine  the stability analysis in our model, we need to determine the second-order actions for scalar and tensor perturbations. In this context, from this action, we can determine whether a cosmological model is free from ghosts and gradient instabilities considering the coefficients $Q_s$ and $c_s$ associated with scalar perturbations. In this respect, the absence of ghosts and Laplacian instabilities for scalar perturbations requires that the coefficients $Q_s>0$ and $c_s^2>0$, respectively. In relation to the tensor perturbations, the no-ghost condition is given by the coefficient $Q_T>0$ associated to the second-order action for the gravitational wave.
However, as we mentioned earlier, the original formulation of Rastall gravity does not arise from a standard variational principle (action formulation).
In the framework of Rastall gravity, the usual conservation law of the energy–momentum tensor is modified according to Eq.(\ref{noncT}) or the generalized expression of the covariant divergence of the energy-momentum tensor given by Eq.(\ref{T1}) and we do not have a Lagrangian formulation.

Nevertheless, even in theories that are not derived from a fundamental action (or Lagrangian), the equations of motion can always be linearized around a homogeneous and isotropic background solution.
Thus, at the perturbative level, it is possible reconstruct an effective second-order action associated to the scalar perturbations $S_s^{(2)}$ that yields the same linearized equations at the perturbative level. An similar situation arises for the second-order action related to the gravitational wave.
In this context, firstly we can expand all dynamical variables  up to second order around the background. After, we need to  insert these expansions into the modified field equations and keeping only first-order terms in the perturbations. Thus, we find a set of coupled equations for the scalar, vector, and tensor degrees of freedom.
In this form, we can obtain  a functional of the perturbations whose Euler–Lagrange equations reproduce exactly those linearized relations. Therefore, 
this functional defines the effective second-order action $S_s^{(2)}$ related with scalar perturbations. A similar methodology can be applied to derive the second-order action for gravitational wave.
This methodology is analogous to what is commonly done in modified gravity theories such as $f(R)$, Horndeski, or scalar–tensor models, among others, see Refs.\cite{DeFelice:2011uc,Kobayashi:2019hrl}.

In this form, following Refs.\cite{DeFelice:2011uc,Kobayashi:2019hrl}  the second-order action for scalar perturbations can be written as
\begin{equation}
S_s^{(2)}=\int\,dt\,d^3x\,a^3\,Q_s\left[\dot{\mathcal{R}}^2-\frac{c_s^2}{a^2}(\partial\mathcal{R})^2\right],
\end{equation}
where $\mathcal{R}$ corresponds to the scalar metric perturbation and the coefficient $Q_s$ associated to the scalar perturbation is defined as
\begin{equation}
Q_s=\frac{X\,p_X}{H^2}=\frac{K_2(\phi)\,X}{H^2},\label{Qs}
\end{equation}
where the coefficient $K_2(\phi)$ is defined by Eq.(\ref{K2}) and the speed of sound is given by Eq.(\ref{cs2}). 
In this way, in the case of the standard Rastall gravity the function $K_2(\phi)$ is a constant and for the situation in which we have the generalized Rastall gravity this function is given by Eq.(\ref{K2}). Thus, in order to avoid the appearance of ghosts and Laplacian instabilities we require the conditions $Q_s>0$ and $c_s^2>0$. For the case of the second-order action of gravitational waves, the coefficient $Q_T$ is defined as $Q_T=M_p^2/4$, see Refs.\cite{DeFelice:2011uc,Kobayashi:2019hrl}. Thus,  the non-ghost condition  $Q_T>0$ for tensor perturbations is automatically satisfied in our models.    In this way, from Eq.(\ref{Qs}), we can note that the coefficient $Q_s$ is always positive, because the parameter $K_2$ defined by Eq.(\ref{K2}) turns out to be $K_2>0$ (see, lower panel of Fig.\ref{fig2} ) and the same applies to $X=\dot{\phi}^2/2$ and $H^2$. In this form, we find that in both Rastall gravities the reconstruction of the  model is free of ghosts, since $Q_s$ together $Q_T$ are coefficients greater than zero. In addition, because the Rastall parameter $\lambda_\text{Ras}$ is a positive quantity, then from Eq.(\ref{cs2}) always the speed of sound $c_s>0$ ensures the absence of Laplacian or gradient instabilities. To confirm these results, we proceed  to  plot these quantities during the inflationary era in both frameworks; standard and generalized Rastall gravity.   

\begin{figure}[ht]
    \centering
\includegraphics[width=0.9\linewidth]{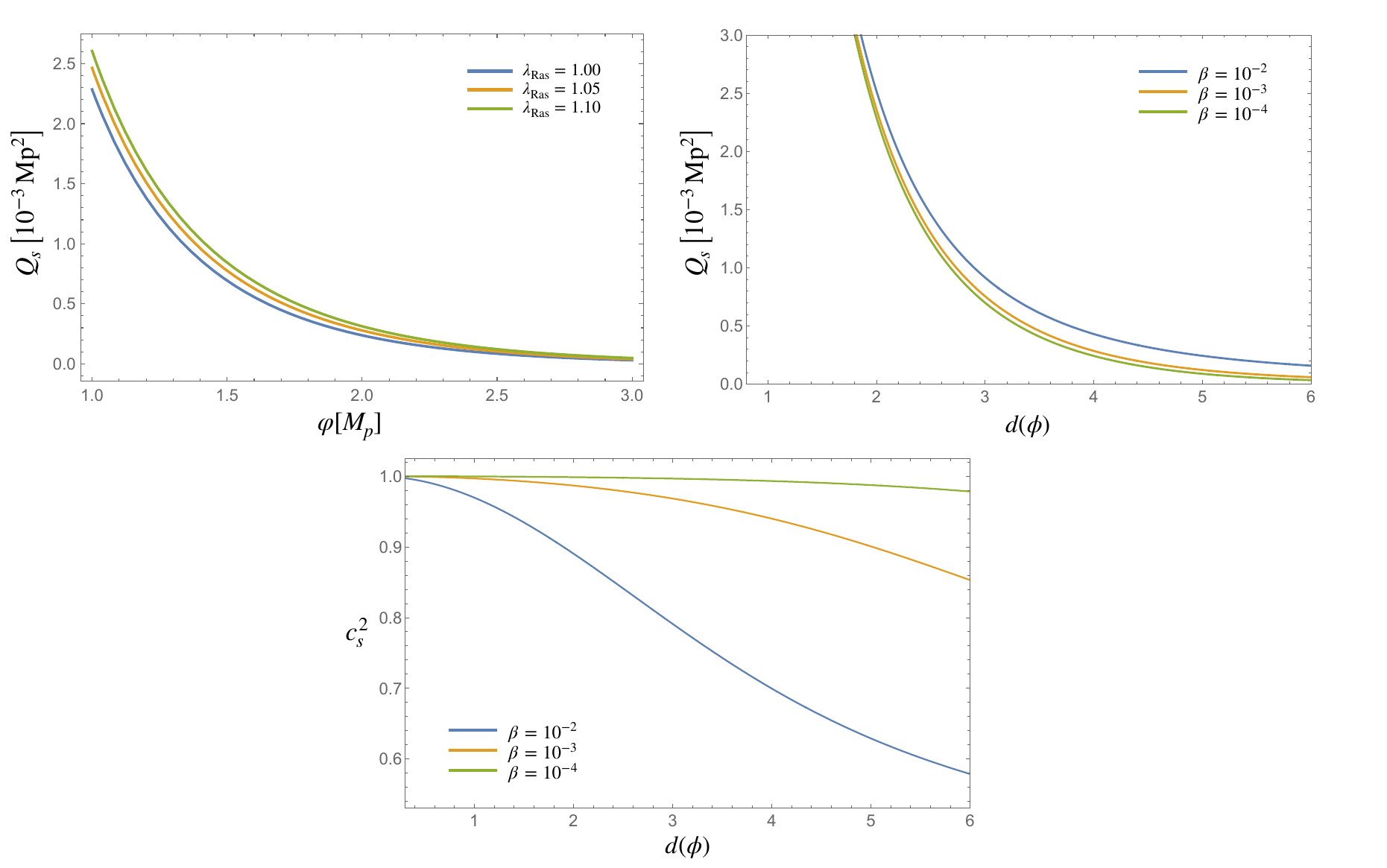}
    \caption{The upper left panel shows the evolution of the coefficient $Q_s$ related to the scalar perturbations in terms of the scalar field $\varphi=\phi-C$, in  standard Rastall gravity for three different constant values  of the parameter $\lambda_\text{Ras}$. The upper right panel presents the evolution of the same coefficient $Q_s$ but in  generalized Rastall gravity as a function of $d(\phi)$ associated to scalar field, for three different values of the parameter $\beta$.  The lower panel displays  the evolution of the speed of sound in generalized Rastall gravity in terms of the function $d(\phi)$ for different values of the parameter $\beta$.  }
    
\label{fig4}
\end{figure}

The Fig.\ref{fig4}, shows the evolution of the coefficient $Q_s$ associated to scalar perturbations  in  standard Rastall gravity (upper left panel) and in  generalized Rastall gravity (upper right panel) in terms of the scalar field. The lower panel displays the behavior 
of the square  speed of sound, $c_s^2$,  as a function of the scalar field in generalized Rastall gravity. 

In relation to the coefficient $Q_s$ associated with the scalar perturbations in  standard Rastall gravity (upper left panel of Fig.\ref{fig4}), where the Rastall parameter $\lambda_\text{Ras}$ is constant, we note that during all the inflationary scenario, this coefficient remains  positive and rather small, of  order  $\mathcal{O}(10^{-3})M_p^2$. Here we have considered three different constant values of the Rastall parameter $\lambda_\text{Ras}$. From this plot, we notice  that as we approach the end of the inflationary period i.e., $\varphi\rightarrow 0$ (see left panel of Fig.\ref{Potencial1}), the coefficient $Q_s$ related to the scalar perturbations grows progressively larger.  
Regarding to the speed of sound in standard Rastall gravity, where the Rastall parameter is a constant,  from Eq.(\ref{19}) we have that this speed always is positive for values of $\lambda_\text{Ras}>2/3$ and in particular for   $\lambda_\text{Ras}\geqslant1$. In this form, we find that in standard Rastall gravity our model avoids the appearance of ghosts and Laplacian instabilities, since the model satisfies the conditions $Q_s>0$, $Q_T>0$ and $c_s^2>0$, respectively. 

Concerning  the coefficient $Q_s$ related with the scalar perturbations in  generalized Rastall gravity (upper right panel of Fig.\ref{fig4}), where the Rastall parameter $\lambda_\text{Ras}$ is variable, we observe that this coefficient $Q_s$  is always positive and rather small, of  order  $\mathcal{O}(10^{-3})M_p^2$ during  most of the inflationary era.  In addition, from this plot we note that as we approach the end of the inflationary period, i.e., as $d(\phi)\rightarrow 0$, the coefficient $Q_s$ increases progressively. In relation to the speed of sound in generalized Rastall gravity, we notice from the lower panel of Fig.\ref{fig4}, that $c_s^2$ is always a  positive quantity.  Also, we note that this speed of sound approaches  one as the inflationary epoch comes to an end, where $d(\phi)\rightarrow 0$. Thus, in the framework of a generalized Rastall gravity, our reconstructed model remains free of ghost and Laplacian instabilities, as the conditions $Q_s>0$, $Q_T>0$ and $c_s^2>0$ 
are fulfilled.

\section{Conclusions }\label{Conc}

In this work, we investigated the reconstruction of inflationary dynamics within the framework of generalized Rastall gravity. Using a general reconstruction formalism, we have derived the effective potential as a function of the scalar field under the slow-roll approximation, guided by cosmological constraints on the scalar spectral index $n_s(N)$ together with the Rastall parameter.

Within this general analysis, we have derived integrable expressions for the effective potential based on the scalar spectral index $n_s(N)$ and the Rastall parameter $\lambda_{\text{Ras}}(N)$, with $N$ denoting the number of $e-$folds. In exploring the reconstruction of inflation in Rastall gravity, we examined two distinct scenarios: the standard case, where the Rastall parameter remains constant, and a generalized extension, in which $\lambda_{\text{Ras}}$ evolves with time.

For the standard Rastall gravity scenario, where the Rastall parameter is taken as a constant, we examined a specific case to reconstruct the effective potential $V(\phi)$, focusing on the attractor relation $n_s - 1 = -2/N$. In this context, we applied the general reconstruction framework, making use of the analytical expressions previously derived. In this scenario, we have found that the potential expressed in terms of the number of $e$-folds coincides with the result obtained in the framework of GR. However, the reconstruction of the effective potential as a function of the scalar field is modified by the inclusion of the Rastall parameter. As in the case of GR, we have found that the inflationary scenario can proceed via two possibilities: starting from positive values of the field $\varphi$ or from negative ones. In both cases, the inflationary epoch ends when the field approaches values close to zero.
In addition, we found that the constraints on the parameters in this scenario are also modified by the presence of the Rastall parameter.

For the generalized Rastall gravity, we have derived a general expression for the reconstruction of the potential in terms of the number of $e$-folds $N$. In this framework, we have obtained a first-order differential equation for the potential $V(N)$. As an example, we have considered the simplest dependence of the Rastall parameter $\lambda_\text{Ras}$ on the number of $e$-folds $N$ (see Eq.(\ref{ec2})) together with the attractor given by Eq.(\ref{ns}) for the scalar spectral index $n_s=n_s(N)$. For this example, we have obtained the analytical reconstruction of the effective potential in terms of the number of $e$-folds $N$ given by Eq.(\ref{VN2}). Here, we have confirmed that in the limit $\beta = 0$, that is, when $\lambda_\text{Ras}$ is constant and equal to 1, the expression of the potential reduces to that obtained in GR. To determine the relation between the number of $e$-folds $N$ and the scalar field, we found that it is not possible to obtain an analytical solution for $N = N(\phi)$. Therefore, to find an analytical solution to Eq.(\ref{ddd}) and subsequently express $N$ as a function of $\phi$, we have assumed that the parameter $\beta$, associated with the Rastall parameter, is small, allowing us to expand the root of the equation. From this approximation, we have obtained the reconstruction of the effective potential and the Rastall parameter in terms of the scalar field given by Eqs.(\ref{VV2}) and (\ref{LL2}), respectively. In addition, we have obtained the different constraints on the parameter-space in this scenario from the observational parameters. Here, we have found that for different fixed values of $\beta$, the constraints obtained on the integration constants $A$ and $B$ are similar. In addition, we have determined the stability analysis (see Section \ref{SA}) of our reconstructed models in both the  standard and generalized Rastall gravity frameworks. Here we have found that in both frameworks these models remain free of ghost and Laplacian instabilities, since  the conditions $Q_s>0$, $Q_T>0$ and $c_s^2>0$ 
are satisfied. 

Finally, in this paper, we have not considered the reconstruction of Rastall gravity for more complex time-dependent (or the number of $e-$folds $N$) functions of the Rastall parameter. As a future step to further strengthen the model's comparison with observational data, a Markov Chain Monte Carlo (MCMC) analysis would provide more robust constraints on the model parameters, especially in  generalized Rastall gravity scenario. We hope to return to these points in the near future.

\bibliography{bio}

\end{document}